\documentstyle[12pt,equation,epsfig]{article}
\begin{document}
\newcommand{\be}{\begin{equation}}
\newcommand{\ben}{\begin{subequations}}
\newcommand{\een}{\end{subequations}}
\newcommand{\beq}{\begin{eqalignno}}
\newcommand{\eeq}{\end{eqalignno}}
\newcommand{\ee}{\end{equation}}
\newcommand{\epem}{\mbox{$e^+ e^-$}}
\newcommand{\tanb}{\mbox{$\tan \! \beta$}}
\newcommand{\mhpl}{\mbox{$m_{H^+}$}}
\newcommand{\tdec}{\mbox{$t \rightarrow H^+ b$}}
\renewcommand{\thefootnote}{\fnsymbol{footnote}}

\begin{flushright}
APCTP 98--11 \\
TIFR/TH/98--11\\
UCRHEP--T228 \\
PRL--TH--98/004 \\
May 1998\\
\end{flushright}
\vspace*{2cm}
\begin{center}
{\Large \bf Light Charged Higgs Bosons in Supersymmetric Models} \\
\vspace*{6mm}
Manuel Drees$^1$, Ernest Ma$^2$, P.N. Pandita$^3$, D.P. Roy$^4$ and
Sudhir K. Vempati$^5$\\
$^1${\it APCTP, 207--43 Cheongryangryi--dong, Tongdaemun--gu, Seoul
130--012, Korea} \\
$^2${\it Physics Department, University of California, Riverside, CA 92521, 
USA} \\
$^3${\it Physics Department, North Eastern Hill Univ., Shillong 793022,
India} \\
$^4${\it Tata Institute of Fundamental Research, Mumbai 400005, India} \\
$^5${\it Theory Group, Physical Research Laboratory, Ahmedabad 380009, India}
\end{center}

\vspace*{1cm}
\begin{abstract}
We point out that present experimental limits from searches for
neutral Higgs bosons at LEP already imply stringent lower bounds on
the mass of the charged Higgs boson in the Minimal Supersymmetric
Standard Model (MSSM); these bounds are especially severe for low
values of \tanb\ ($\tanb \leq 3$), where the $H^+ \bar{t} b$ coupling
is large. However, these indirect constraints are much weaker in
simple extensions of the MSSM Higgs sector involving the introduction
of an extra $U(1)$ gauge group or an extra $SU(2) \times U(1)_Y$ Higgs
singlet field; in the latter case charged Higgs bosons can even be
light enough to be pair produced at LEP.

\end{abstract}
\clearpage
\setcounter{page}{1}

The Higgs mechanism offers the theoretically best understood
description of electroweak symmetry breaking, which is required to
generate masses for $W$ and $Z$ gauge bosons as well as matter
fermions (quarks and leptons). The minimal Standard Model (SM)
predicts the existence of only a single physical Higgs boson, a
neutral CP--even particle. However, in many extensions of the SM the
Higgs sector is more complicated. In particular, in models with two or
more Higgs doublets the spectrum of physical Higgs fields contains
some charged Higgs bosons.

The currently best motivated extension of the SM involves the introduction
of softly broken supersymmetry, which stabilizes the gauge hierarchy
against radiative corrections \cite{2a}. Besides predicting the existence
of superpartners of all known particles, realistic supersymmetric theories
also contain \cite{2} at least two Higgs doublet superfields, to allow
for anomaly cancellation in the higgsino sector and to give masses to both
hypercharge $Y= +1/2$ and $Y = -1/2$ matter fermions. This second Higgs
doublet superfield also plays a crucial role in the unification of all
known gauge interactions, which is natural in supersymmetric extensions
of the SM, but leads to conflict with LEP data in the SM itself
\cite{3}. Finally, loops involving superpartners allow to cancel
\cite{4} potentially large, positive contributions from $t - H^+$ loops 
to the partial width for radiative $b \rightarrow s \gamma$ decays,
thereby avoiding stringent lower bounds \cite{5} on the mass of
charged Higgs bosons in non--supersymmetric models. Supersymmetry therefore
now appears to be an (almost) necessary condition for the existence of
charged Higgs bosons that are light enough to be produced in the decay
of top quarks.

On the other hand, unsuccessful searches for neutral Higgs bosons at
LEP have now reached a sensitivity that begins to impose nontrivial
constraints on charged Higgs bosons in many supersymmetric models. The
reason is that at least the simplest, and hence most attractive,
supersymmetric models contain fewer free parameters in the Higgs
sector than a general (non--supersymmetric) model with two Higgs
doublets does. The purpose of this note is to explore these {\em
indirect} lower bounds on the mass of the charged Higgs boson
quantitatively, within the framework of three different supersymmetric
models.

The first of these models is the minimal supersymmetric standard model
(MSSM), which is a straightforward supersymmetrization of the SM. In
particular, the gauge group is $SU(3) \times SU(2) \times U(1)_Y$, and
the Higgs sector \cite{6} consists of only two Higgs doublet superfields.
The physical spectrum of Higgs bosons therefore contains five fields:
Two neutral CP--even fields $h, H$; one neutral CP--odd field $A$;
and charged Higgs bosons $H^\pm$. At the tree level the masses and
interactions of these Higgs bosons are determined by just two free
parameters, e.g. the mass $m_A$ of the CP--odd state and \tanb, the ratio 
of the vacuum expectation values (vev) of the neutral components of
the two Higgs doublets.

However, the masses and couplings of the neutral CP--even states
receive potentially large radiative corrections \cite{7} from loops
involving top quarks and their spin--0 superpartners, the stops. We
treat these one--loop corrections using the effective potential method
\cite{8}.  As pointed out in refs.\cite{9}, one can absorb the
dominant two--loop QCD corrections by using a running ($\overline{\rm
MS}$ or $\overline{\rm DR}$) top mass at appropriately chosen scale in
the $m_t^4$ factors appearing in the expressions for the one--loop
corrections. We take a pole top mass of 175 GeV, which corresponds to
$m_t(m_t) \simeq 166$ GeV and $m_t(1 \ {\rm TeV}) \simeq 151$ GeV.
The size of these corrections increases logarithmically with the stop
mass scale. We therefore conservatively take 1 TeV for the soft
breaking masses of both the $SU(2)$ doublet and $SU(2)$ singlet stops
$\tilde{t}_L$ and $\tilde{t}_R$, the largest value commonly accepted as
being compatible with naturalness arguments. We also allow for mixing
between the two stop current eigenstates. This can increase the
corrections significantly, with $m_h$ becoming maximal \cite{10} if
$A_t + \mu \cot \! \beta = \sqrt{6} m_{\tilde t}$, where $A_t$ is a
trilinear soft breaking parameter, $\mu$ the supersymmetric higgsino
mass parameter, and $m_{\tilde t}$ the common $\tilde{t}_L$ and
$\tilde{t}_R$ soft breaking mass; we will refer to this choice of stop
mixing parameters as ``maximal mixing''.

The model remains fairly constrained even after including these
radiative corrections. In particular, the mass of the lightest neutral
scalar is bounded from above:
\be \label{e1}
m_h^2 \leq M_Z^2 \cos^2(2\beta) + \epsilon(m_t, m_{\tilde t}, A_t),
\ee
where $\epsilon$ parameterizes the effect of the radiative corrections
described above. Note that $\epsilon$ is approximately independent of
\tanb; for large $m_A$, $m_t=175$ GeV and $m_{\tilde t}=1$ TeV it
amounts to about $0.9 M_W^2 \ (1.6 M_W^2)$ for no (maximal) stop mixing.
It is important to note that the bound (\ref{e1}) can {\em only} be
saturated for large $m_A$. In the region of small\footnote{We will
always assume $\tanb > 1$ here; this is required for the top Yukawa
coupling to remain perturbative up to some high energy scale, and more
generally seems indicated by the large ratio $m_t/m_b$.} \tanb\ the
current LEP limits \cite{11} on the masses of neutral Higgs bosons
therefore already imply a quite stringent lower bound on $m_A$. This in
turn constrains the mass of the charged Higgs boson, which is given by
\be \label{e2}
m^2_{H^+} = M_W^2 + m_A^2 + \epsilon_+,
\ee
where the radiative correction $\epsilon_+$ is small and can be of either
sign \cite{9}.

This is illustrated by the dotted curves in Fig.~1, which show the
indirect lower bound on \mhpl\ that follows from LEP searches for
neutral MSSM Higgs bosons; the upper (lower) curve is for no (maximal)
stop mixing. The LEP search limits have been interpreted as implying
the following constraints on tree--level Higgs production cross sections:
\ben \label{e3} \beq
\sigma(\epem \rightarrow Z h) &< 0.35 \ {\rm pb}; \label{e3a} \\
\sigma(\epem \rightarrow h A) &< 0.12 \ {\rm pb}, \label{e3b}
\eeq \een
at center--of--mass energy $\sqrt{s}=183$ GeV. The bound (\ref{e3a})
corresponds to a lower limit of 88.7 GeV on the mass of the SM Higgs
boson, while (\ref{e3b}) implies $m_h \simeq m_A \geq 75$ GeV for
$\tanb \gg 1$ in the MSSM. We have not attempted to combine the searches
for $Zh$ and $hA$ production, since these final states have different
backgrounds. The lower bound $m_A > 75$ GeV, which follows from
(\ref{e3b}), implies $\mhpl > 109$ GeV; this explains the flat parts in
the dotted curves in Fig.~1. 

However, for low \tanb, the constraint (\ref{e3a}) gives a stronger
bound on $m_A$, and hence on \mhpl. This has important ramifications
for charged Higgs searches at hadron colliders such as the
Tevatron. The most promising searches \cite{12} all rely on \tdec\
decays. Since the relevant $H^+ \bar{t} b$ couplings are proportional
to $m_t \cot \! \beta \pm m_b \tanb$, for a given value of \mhpl\ the
branching ratio for such decays is large at small and at large \tanb,
but has a pronounced minimum at $\tanb \simeq \sqrt{m_t/m_b} \simeq
7.5$. Fig.~1 shows that already now LEP searches for neutral Higgs
bosons imply that in the MSSM \tdec\ decays are possible only for
$\tanb > 2.3 \ (1.4)$ for no (maximal) stop mixing.

We emphasize, however, that these constraints are entirely {\em indirect},
stemming from the search for neutral Higgs bosons. This motivated us to
investigate the question to what extent these lower bounds on \mhpl\
can be relaxed by only modifying the neutral Higgs sector, keeping the
charged Higgs sector unchanged. This implies that we restrict ourselves
to models containing additional $SU(2)$ singlet Higgs superfields and/or
some new interactions.

To be specific, we studied two fairly modest extensions of the MSSM Higgs
sector. The first of these models is a specific realization of the
so--called superstring--inspired $E(6)$ models \cite{15}. In general the
Higgs sector of even the simplest such models \cite{16} differs quite 
substantially from that of the MSSM. However, as pointed out in 
refs.\cite{17}, under certain assumptions one ends up with models that 
contain only one more parameter in the Higgs sector than the MSSM does.
The first assumption is that the mass of the new neutral $Z'$ gauge
boson present in these models is large compared to $M_Z$. In most
cases current $Z'$ mass limits are in fact already so large \cite{18}
that this condition is automatically satisfied. This implies that the
vev of the new $SU(2) \times U(1)_Y$ singlet Higgs field $N$ must be
much larger than the vevs of the $SU(2)$ doublets. The second
assumption is that the trilinear soft breaking term associated with
the $N H_1 H_2$ term in the superpotential is {\em not} large; here
$H_1$ and $H_2$ are the $Y=-1/2$ and $Y=+1/2$ Higgs doublets, 
respectively. Under these assumptions the singlet Higgs field $N$ is
much heavier than the doublets, and does not mix with them. However,
the trilinear scalar $N H_1 H_2$ interaction gets a large supersymmetric
contribution $\propto \langle N \rangle$. Some $N-$exchange contributions
to quartic Higgs couplings therefore remain even after $N$ is integrated
out. Furthermore, the existence of a new $U(1)$ factor leads to new
$D-$term contributions to the Higgs potential.

We refer the reader to refs.\cite{17} for further details of these
models. Here we merely state that the upper bound (\ref{e1}) on the
mass of the lightest neutral CP--even state $h$ gets modified to
\be \label{e5}
m^2_h \leq M_Z^2 \cos^2(2 \beta) + \frac {\lambda^2} {\sqrt{2} G_F}
\left[ \frac{3}{2} + (2a-1) \cos (2 \beta) - \frac{1}{2}
\cos^2(2\beta) - \frac {\lambda^2} {g_x^2} \right] + \epsilon.
\ee
Here, the radiative correction $\epsilon$ is the same as in eq.(\ref{e1}),
$G_F$ is the Fermi constant, the constant $a$ depends on the $E(6)$
symmetry breaking pattern, $\lambda$ (called $f$ in refs.\cite{17})
is the $N H_1 H_2$ superpotential coupling, and the coupling $g_x$
associated with the extra $U(1)$ can in most models be set equal to
the standard hypercharge coupling $g_1$. As advertised, a given model
only introduces a single new parameter $\lambda$. As an illustration we
consider the so--called $\eta-$model, where $a=0.2$. In this case
consistency of the model requires \cite{17} $\lambda \leq 0.35$.

The new, positive contribution in eq.(\ref{e5}) makes it easier to
satisfy the Higgs search constraints from LEP, allowing for a reduced
value of $m_A$ compared to the MSSM. Moreover, the relation between
$m_A$ and \mhpl\ also gets modified:
\be \label{e6}
m^2_{H^+} = m_A^2 + M_W^2 \left( 1 - \frac {2 \lambda^2} {g_2^2}
\right) + \epsilon_+,
\ee
where $g_2$ is the $SU(2)$ gauge coupling. This further reduces the lower
bound on \mhpl. Note that the new contributions to eqs.(\ref{e5}) and
(\ref{e6}) are maximized for different values of $\lambda$. The
reduction of \mhpl\ for fixed $m_A$ is obviously maximal for the largest
allowed value of $\lambda$, 0.35 for the $\eta-$model, while the new
contribution to eq.(\ref{e5}) is maximal for $\lambda \simeq 0.27$ (for
$a=0.2, \ g_x = g_1 \simeq 0.35$). A numerical scan of the parameter
space reveals that the absolute minimum of \mhpl\ is reached for
$\lambda \simeq 0.27$ if $\tanb \leq 2$, and for $\lambda = 
\lambda_{\rm max} = 0.35$ for $\tanb > 2.5$.

The results of this scan are shown by the solid curves in Fig. 1; the
upper (lower) curve again refers to no (maximal) stop mixing. Since
this model contains exactly the same (potentially) light Higgs fields
as the MSSM, the constraints (\ref{e3}) can be applied without any
modification; in particular, for large \tanb\ the CP--odd state is
again nearly degenerate with one of the CP--even Higgs bosons.

In the region of small \tanb\ the effects of this rather modest
modification of the MSSM Higgs sector are quite dramatic.  In
particular, \tdec\ decays are now again allowed all the way down to
$\tanb = 1.3$ even in the absence of stop mixing. If stop mixing is
maximal, such decays are possible even for $\tanb = 1.0$; however, for
such a low value of \tanb\ the top Yukawa coupling would have a Landau
pole at an energy scale quite close to the weak scale. The
modification of the lower bound on \mhpl\ is more modest for $\tanb
\geq 4$, where the limit (\ref{e3b}) provides the most stringent
constraint. The reason is that the resulting bound on $m_A$ is
essentially the same as in the MSSM, so the difference between the two
bounds is entirely due to the new contribution to the charged Higgs
mass in eq.(\ref{e6}).

The third model we investigate is the so--called next--to--minimal
supersymmetric standard model (NMSSM). It differs from the MSSM only
in the Higgs sector, where one postulates \cite{19} the existence of
an $SU(2) \times U(1)_Y$ singlet superfield $N$. The model is
therefore conceptually far simpler than the $E(6)$ models discussed
above; nevertheless the modification of the Higgs sector is more
extensive.  Even if we restrict ourselves to purely cubic terms in the
superpotential $f$, gauge symmetry allows one to introduce two
different Higgs self--couplings:
\be \label{e7}
f_{\rm Higgs} = \lambda N H_1 H_2 - \frac {k}{3} N^3,
\ee
where we have used the notation of ref.\cite{20}. Together with the
corresponding soft breaking terms, there are six free parameters in the
Higgs sector, even after we fix the sum of the squares of the vevs
of the $SU(2)$ doublets to reproduce the known mass of the $Z$ boson.
Moreover, the spectrum now contains three neutral CP--even fields
$H_i$ and two CP--odd fields $A_i$ in addition to the charged Higgs field
$H^\pm$.

Nevertheless one can still derive \cite{21} an upper bound on the mass
of the lightest scalar Higgs field. After including radiative 
corrections, one has \cite{22,20}:
\be \label{e8}
m^2_{H_1} \leq M_Z^2 \cos^2(2\beta) + \frac {2 \lambda^2 M_W^2} {g_2^2}
\sin^2(2\beta) + \epsilon,
\ee
where $\epsilon$ is again the same as in eq.(\ref{e1}). Clearly this bound
is only useful if an upper limit for $\lambda$ can be found. Such a limit
can be derived \cite{21,20} from the requirement that all couplings of
the model remain in the perturbative regime up to some very high energy
scale, usually taken to be of the order of the GUT scale.

In Fig.~2 we show the resulting upper bound on $\lambda$ as a function 
of the value of the top Yukawa coupling $h_t$ at scale $M_Z$. We have
used two--loop renormalization group equations \cite{20} to derive
this bound, with $\alpha_s(M_Z) = 0.120$. We have conservatively taken a 
rather low input scale ($M_Z$, rather than $m_t$) and a rather high
value for the GUT scale ($3 \cdot 10^{16}$ GeV); the resulting bound only
depends weakly on these scale choices. In the absence of
sparticle loop corrections, the top Yukawa coupling is given by
\be \label{e9}
h_t = \frac {g_2 m_t} { \sqrt{2} M_W \sin \! \beta},
\ee
where $m_t$ is the running top mass; this gives $h_t(m_t) \geq 0.96$
for $m_t({\rm pole})=175$ GeV.\footnote{In general there can be substantial
stop--gluino loop corrections to eq.(\ref{e9}) \cite{23}. However, these
will be small for large values of the stop masses, which maximize the
radiative correction $\epsilon$.} The upper branch of the curve in Fig.~2
is determined by the requirement that $\lambda$ remains in the perturbative
regime, while the sharp drop--off to the right comes from the requirement
that $h_t$ remains perturbative. The absolute upper bound on $h_t$
corresponds to the well--known ``fixed point'' solution \cite{24}; of
particular interest to us is the corresponding lower bound on \tanb,
which can be written in the form
\be \label{e10}
\sin \! \beta \geq 0.84 \frac {m_t(\rm pole)} {175 \ {\rm GeV} }.
\ee
We remark that the bound (\ref{e10}) also applies to the MSSM if one
requires $h_t$ to remain perturbative up to scale $M_X = 3 \cdot 10^{16}$
GeV; we have extended the dotted curves in Fig.~1 to lower values of
\tanb\ in order to allow for possible intermediate scales, which could
relax this bound \cite{24a}.

The relation between the masses of charged and neutral CP--odd Higgs
bosons also gets modified in the NMSSM \cite{21,25}:
\be \label{e11}
m^2_{H^+} = M_W^2 \left( 1 - \frac {2 \lambda^2} {g_2^2} \right)
+ m^2_{A'} + \epsilon_+,
\ee
where $m^2_{A'}$ is the mass of the neutral $SU(2)$ doublet CP--odd
state in the absence of doublet--singlet mixing. Note that this mixing
can only reduce the mass of the lighter CP--odd state, i.e.
$m^2_{A'} \geq m^2_{A_1} \geq 0$. On the other hand, the $\lambda^2-$term
in eq.(\ref{e11}) obviously reduces the mass of the charged Higgs boson.
Moreover, doublet--singlet mixing can also reduce the couplings of the light
physical Higgs states to gauge bosons \cite{26}; the bounds on $m_{H_1}$
and $m_{A_1}$ in the NMSSM are therefore much weaker \cite{27,25} than
those on $m_h$ and $m_A$ in the MSSM.

We have interpreted these experimental constraints as follows. The LEP2
search limits (\ref{e3}) were taken to limit the sums $\sum_i
\sigma(\epem \rightarrow Z H_i)$ and $\sum_{i,j} \sigma(\epem \rightarrow
H_i A_j)$, respectively. However, unlike in the MSSM these bounds from
searches at the highest available center--of--mass energy did not
supersede the older LEP1 constraints completely, since the LEP2 constraints
allow very light $H_1, \ A_1$ if they are dominantly $SU(2)$
singlets. It turns out that the LEP1 constraints on the couplings of
such light states to $Z$ bosons are often stronger than those from
higher energies. We have parametrized the ALEPH Higgs search limits
from their lower energy data \cite{28} as follows:
\ben \label{e12} \beq
\left( g_{Z Z H_i} \right)^2 &\leq \left( \frac {g_2 M_Z} 
{\cos \! \theta_W} \right)^2 \ \cdot \ 1.6 \cdot 10^{-4}
e^{m_{H_i}/8.2} ; 
\label{e12a} \\
\left( g_{Z H_i A_j} \right)^2 &\leq \left( \frac {g_2} {2 \cos \! \theta_W}
\right)^2 \cdot \left\{ 
\mbox{$ \begin{array}{lr}
0.1, & m_{H_i} + m_{A_j} \leq 81 \ {\rm GeV} \\
\left[ 0.1 ( m_{H_i} + m_{A_j} ) - 8.0 \right], & 
m_{H_i} + m_{A_j} > 81 \ {\rm GeV} 
\end{array} $}, \right.
\label{e12b}
\eeq \een
where all masses are in GeV. Note that eqs.(\ref{e12}) apply to individual 
couplings; we have not attempted any summation over related states,
unlike in our NMSSM modification of the LEP2 bounds (\ref{e3}). However,
we have checked that such a summation would not affect the derived
lower limit on \mhpl\ significantly.

This limit is shown by the dashed curves in Fig. 1. Note that this
bound shows little sensitivity to stop mixing even in the low \tanb\
region, unless \tanb\ lies just above the lower bound (\ref{e10}).
More importantly, for $1.7 \leq \tanb \leq 2.5$ the direct $H^+$ search
limit from DELPHI \cite{29}, $\mhpl \geq 53$ GeV, can be saturated in
the NMSSM, in sharp contrast to the other two models. We should mention
that the indirect lower bound on \mhpl\ is not only determined by
the upper bound on $\lambda$ shown in Fig.~2 and the various LEP search
limits described above, but also by the requirement that the desired
minimum of the Higgs potential, where all three neutral Higgs fields 
have non-vanishing vev,\footnote{We need non-vanishing $ \langle H_1^0
\rangle$ and $\langle H_2^0 \rangle$ to give masses to all matter 
fermions. This then automatically implies $\langle N \rangle \neq 0$
at stationary points of the potential.} is the {\em absolute} minimum
of the potential. In particular, the allowed parameter space is 
constrained significantly by requiring that solutions where only
$H_2^0$ or only $N$ have non-vanishing vev should {\em not} be the
absolute minimum.

The interplay of these constraints makes it difficult to give an
analytical explanation for the behavior of the dashed curves in Fig.~1.
It is clear from eq.(\ref{e8}), however, that the upper bound on
$m_{H_1}$ becomes independent of $\lambda$ for large \tanb, where
it approaches the MSSM value. Our numerical scan of the parameter
space finds that for $\tanb \leq 3.5$, \mhpl\ takes its smallest possible
value if $\lambda$ is at its maximum. The presence of intermediate
scales, which could increase the upper bound on $\lambda$ \cite{24a},
can therefore reduce the lower bound on \mhpl\ even further in this
region. Moreover, it would allow for a relative low \mhpl\
even below the lower limit (\ref{e10}) on \tanb.
However, for $\tanb > 4$, \mhpl\ is minimized for $\lambda$ 
around 0.4. This optimal value of $\lambda$ is coincidentally quite
close to the maximal allowed $\lambda$ in the $E(6) \ \eta-$model. As a
result, for $\tanb > 4$ the indirect lower bound on \mhpl\ that
can be derived from neutral Higgs searches at LEP is quite close
for these two models.\footnote{In this region \mhpl\ is minimized in
the NMSSM if the coupling $k$ takes its maximum value. However, the
$k-$dependence of the bound is quite mild. We therefore simply
require $|k| \leq 0.5$ for small and moderate values of $\lambda$, in
agreement with results of refs.\cite{21,25}.}

We have also used our program to search for NMSSM parameters that
allow the decay chain $\tdec \rightarrow W^+ (H_1, A_1) b$. The main
signature for such decays would resemble that for $H^+ \rightarrow W^+
b \bar{b}$ three--body decays \cite{30}, except that there would be a
peak in the $b \bar{b}$ invariant mass spectrum. The light neutral
Higgs boson could also decay into $\tau^+ \tau^-$ pairs, with
branching ratio of order 10\%. We found that in the NMSSM such
scenarios can indeed be realized for small values of \tanb. For
example, for $\mhpl=150$ GeV and $\tanb=1.6$, we found that the
partial widths for $H^+ \rightarrow H_1 W^+ \ (A_1 W^+)$ can exceed
the sum of $H^+ \rightarrow c \bar s$ and $H^+ \rightarrow \tau^+
\nu_\tau$ partial widths by a factor of more than 7.5 (150). Light
CP--odd states can be produced more copiously in $H^+$ decays, since
they cannot be produced singly at LEP, unlike neutral CP--even states;
hence they can have much larger $SU(2)$ doublet components than
CP--even states with the same mass.  Such ``unusual'' $H^+$ decays
allow one to evade \cite{30} bounds on \tdec\ decays based on either
direct searches for enhanced $\tau$ production, or on the reduction of
$t \bar t$ events containing one or two hard leptons (electrons or
muons) \cite{31}.

Finally, we have attempted to assess the impact of future searches for
neutral Higgs bosons at LEP on the lower bound on \mhpl. It now seems
that the ultimate energy of LEP will be around $\sqrt s = 200$ GeV
\cite{moriond}.  Using results of the LEP2 Higgs working group
\cite{lepwg} we estimate that this could give lower limits of 107 GeV
for an SM--like neutral Higgs boson, and of 93 GeV for degenerate
CP--even and CP--odd states with full coupling to the $Z$ (as in the
MSSM at large \tanb). These bounds assume that no indication of a
signal is found; the discovery reach of LEP operating at this energy
would be a few GeV lower. These possible future constraints can be
implemented by requiring
\ben \label{en1} \beq
\sigma(\epem \rightarrow Z h) &< 0.16 \ {\rm pb}; \label{en1a} \\
\sigma(\epem \rightarrow h A) &< 0.025 \ {\rm pb}, \label{en1b}
\eeq \een
at center--of--mass energy $\sqrt{s}=200$ GeV. In case of the NMSSM,
we have again summed over $Z H_i$ and $H_i A_j$ final states when applying
these constraints.

The resulting bounds are shown in Fig.~3, using the same notation as in 
Fig~1. In the MSSM, a failure to detect neutral Higgs bosons at LEP
would lead to the absolute lower bounds $\tanb > 3 \ (1.5)$ for no (maximal)
stop mixing; charged Higgs bosons would then be accessible to top 
decays only for $\tanb > 4.75$ (2.6). Even in the $U(1)_\eta$ model
a nontrivial lower bound on \tanb\ would emerge unless stop mixing is
substantial. On the other hand, in the NMSSM the charged Higgs boson
could still be light enough to be produced in top decays even very close
to the lower bound (\ref{e10}) on \tanb. However, even in this model
a nontrivial absolute lower bound on \mhpl\ of about 80 GeV could be
derived. This is only slightly lower than the ultimate sensitivity of
LEP for direct charged Higgs searches; in the absence of a signal for
neutral Higgs boson production, searches for charged Higgs bosons at LEP
could give additional constraints on parameter space only for
$1.7 \leq \tanb \leq 2.7$. 

Finally, all three models again give fairly similar indirect lower
limits on \mhpl\ for $\tanb \geq 7$.  Indeed the ultimate LEP lower
limit of 110--120 GeV on \mhpl\ (Fig.~3) would hold throughout the
region $\tanb \geq7$. Thus there is plenty of room for direct $H^+$
search via \tdec\ decays in this region even in the MSSM.  It may be
noted here that a direct $H^+$ search via \tdec\ with the Tevatron
collider data has recently given a lower limit of 100--120 GeV on
\mhpl\ for $\tanb = 40-50$ \cite{33}. This result holds in the MSSM as
well as its extensions discussed here. The main difference between
them lies in the relatively low \tanb\ region, where the negative
result from neutral Higgs boson search at LEP implies a severe lower
limit on \mhpl\ in the former case but not the latter.

In summary, we have pointed out that experimental searches at LEP
already impose significant indirect lower bounds on the
mass of the charged Higgs boson in the MSSM. These bounds are particularly
severe for low \tanb. This region is of special interest since here
the $H^+ \bar t b$ coupling is large, while the partial widths of $H^+$
into light SM fermions is small, allowing other interesting decay
modes of the charged Higgs boson to have sizable branching ratios.

However, we found that these indirect lower bounds on \mhpl\ can be
relaxed considerably in relatively modest extensions of the
MSSM. Specifically, an extension based on $E(6)$ models with an extra
$U(1)$ factor, which only adds one new parameter to the description of
the low--energy Higgs sector, is sufficient to re--introduce the
possibility of a substantial branching fraction for \tdec\ decays,
although in these models the smallest allowed value of \mhpl\ still
lies beyond the region that can be covered by searches for $H^+ H^-$
production at LEP. An even more dramatic reduction of the indirect
lower bound on \mhpl\ becomes possible in the NMSSM, where one adds
one $SU(2) \times U(1)_Y$ singlet Higgs superfield to the MSSM.  In
this model the charged Higgs boson could still be light enough to be
discovered at LEP, if \tanb\ lies between 1.7 and 4, or even lower if
one allows for intermediate scales. In this range of \tanb\ the
charged Higgs boson might dominantly decay into an on--shell $W$ boson
and a light neutral Higgs boson, which would complicate the search for
\tdec\ decays at hadron colliders.  On the other hand, we found the
indirect lower bound on \mhpl\ to be less model--dependent for
intermediate and large \tanb. For $\tanb > 4$ the bound in the $E(6)$
model or the NMSSM lies within 10 to 15 GeV of the MSSM value, which
is itself quite modest.

If LEP experiments fail to observe a signal for neutral Higgs boson
production after accumulating several hundred pb$^{-1}$ of data
at $\sqrt{s} = 200$ GeV, even in the $E(6)$ models $\tdec$ decays
would become impossible for $\tanb \leq 2$, unless there is
substantial mixing in the stop sector. However, even in this
pessimistic scenario light Higgs bosons at small $\tanb$ could still
be accommodated in the NMSSM, independent of stop mixing. We conclude
that there remains plenty of parameter space for light charged Higgs
bosons at small and moderate values of \tanb.

\subsection*{Acknowledgements}
We thank the organizers of the Fifth Workshop on Particle Physics
Phenomenology (WHEPP5), where this project was initiated. We thank
Anjan Joshipura for discussions. EM was supported in part by the
US Department of Energy under grant no. DE--FG03--94ER40837,
while PNP was supported in part by the Indian 
Department of Science and Technology under project
No. SP/S2/K--17/94.

\clearpage
\noindent
\setcounter{figure}{0}

\begin{figure}[htb]
\centerline{\epsfig{file=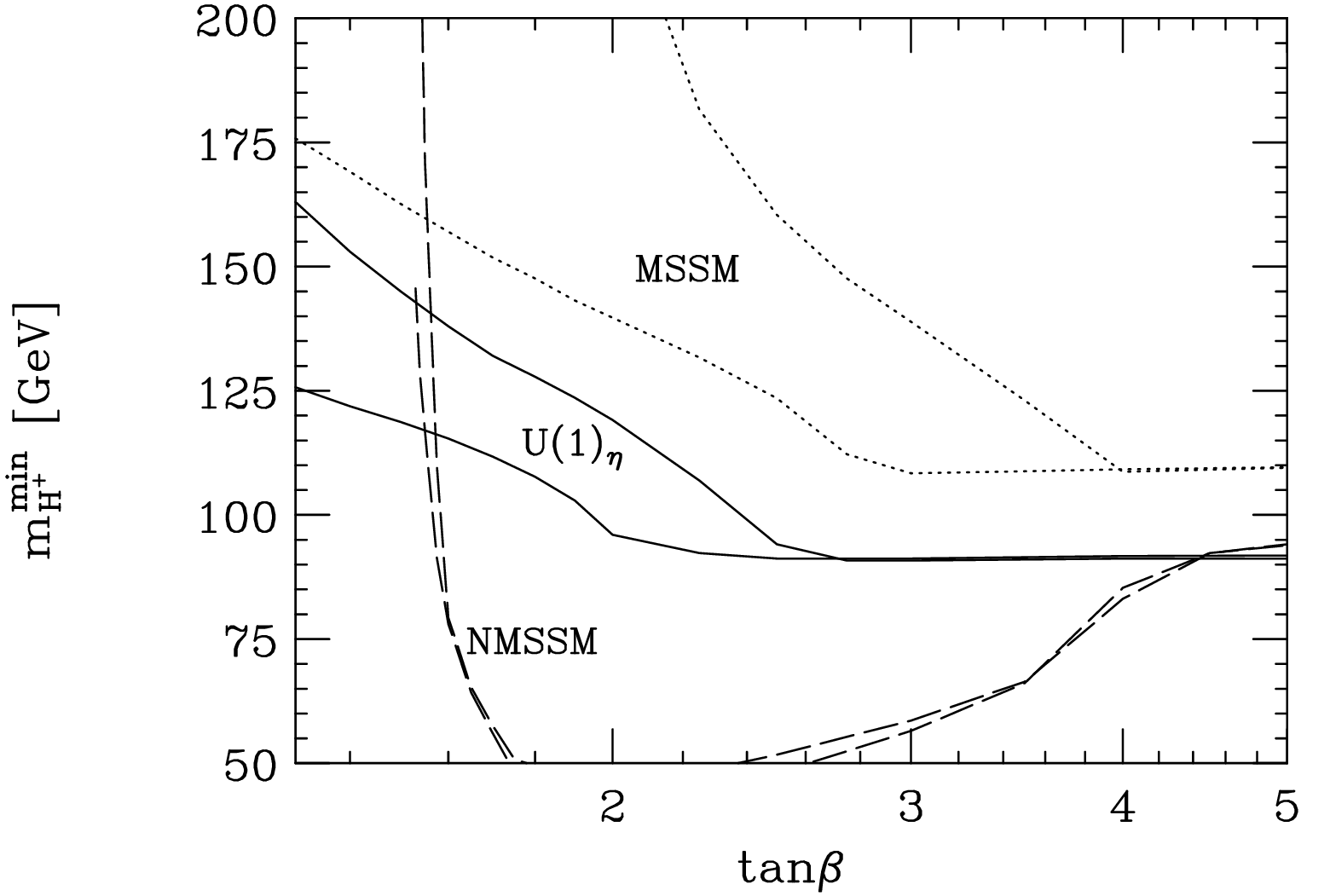,height=11cm}}

\caption
{The indirect lower bound on the mass of the charged Higgs boson that
follows from the searches for neutral Higgs bosons at LEP. The dotted
curves are for the MSSM, the solid ones for the $E(6) \ \eta-$model,
and the dashed ones for the NMSSM. Radiative corrections to Higgs mass 
matrices have been included using $m_t({\rm pole}) = 175$
GeV and a common stop soft breaking mass of 1 TeV; the upper (lower) curve
of a given pattern is for no (maximal) mixing between $SU(2)$ doublet
and singlet stops, as described in the text.}
\end{figure}

\clearpage

\begin{figure}[htb]
\centerline{\epsfig{file=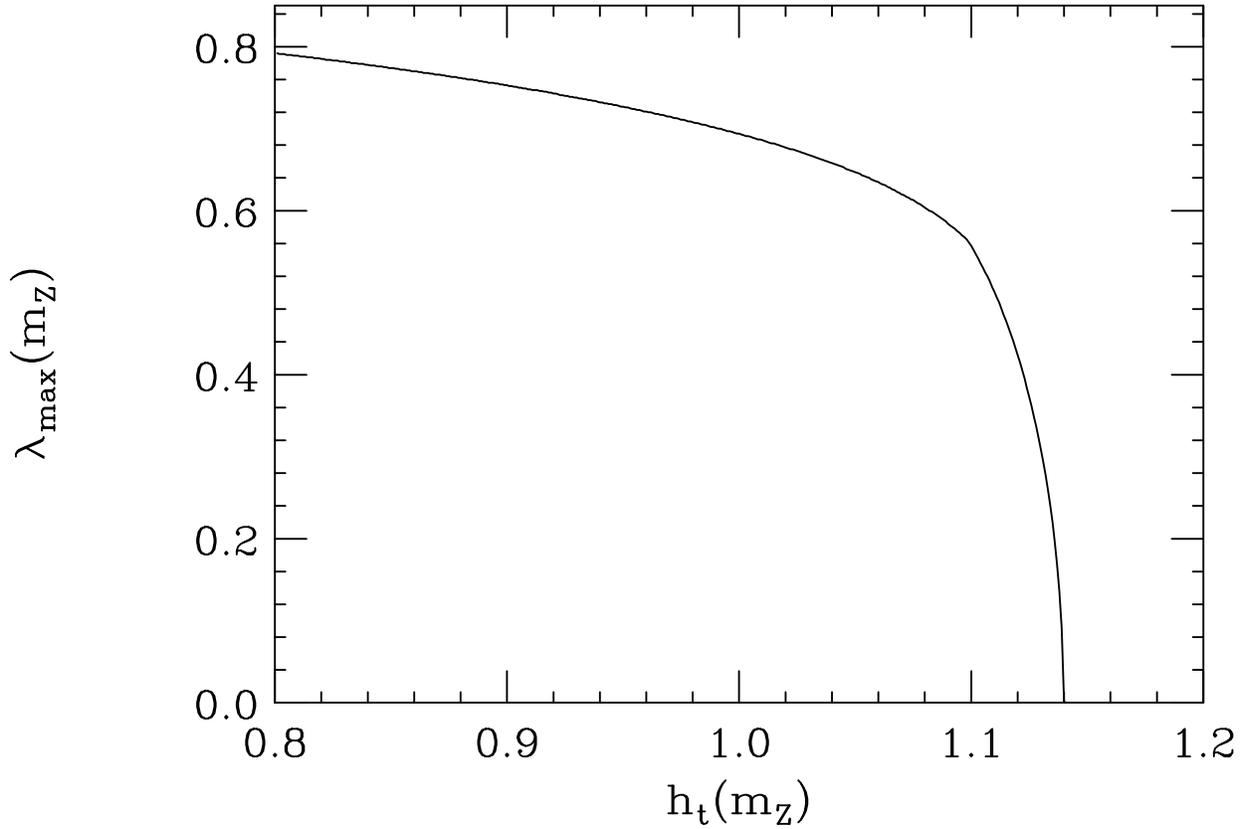,height=11cm}}

\caption
{The upper bound on the Higgs self coupling $\lambda$ that follows from
the requirement that all couplings remain in the perturbative regime
up to scale $M_X = 3 \cdot 10^{16}$ GeV is shown as a function of the
top Yukawa coupling $h_t$. This curve is valid if all other superpotential
couplings are small. The bound on $\lambda$ therefore becomes more
stringent if $\tan \! \beta$ is very large, in which case the bottom
Yukawa coupling becomes sizable. We have assumed $\alpha_s(M_Z) = 0.12$.}

\end{figure}

\clearpage

\vspace*{2cm}

\begin{figure}[htb]
\centerline{\epsfig{file=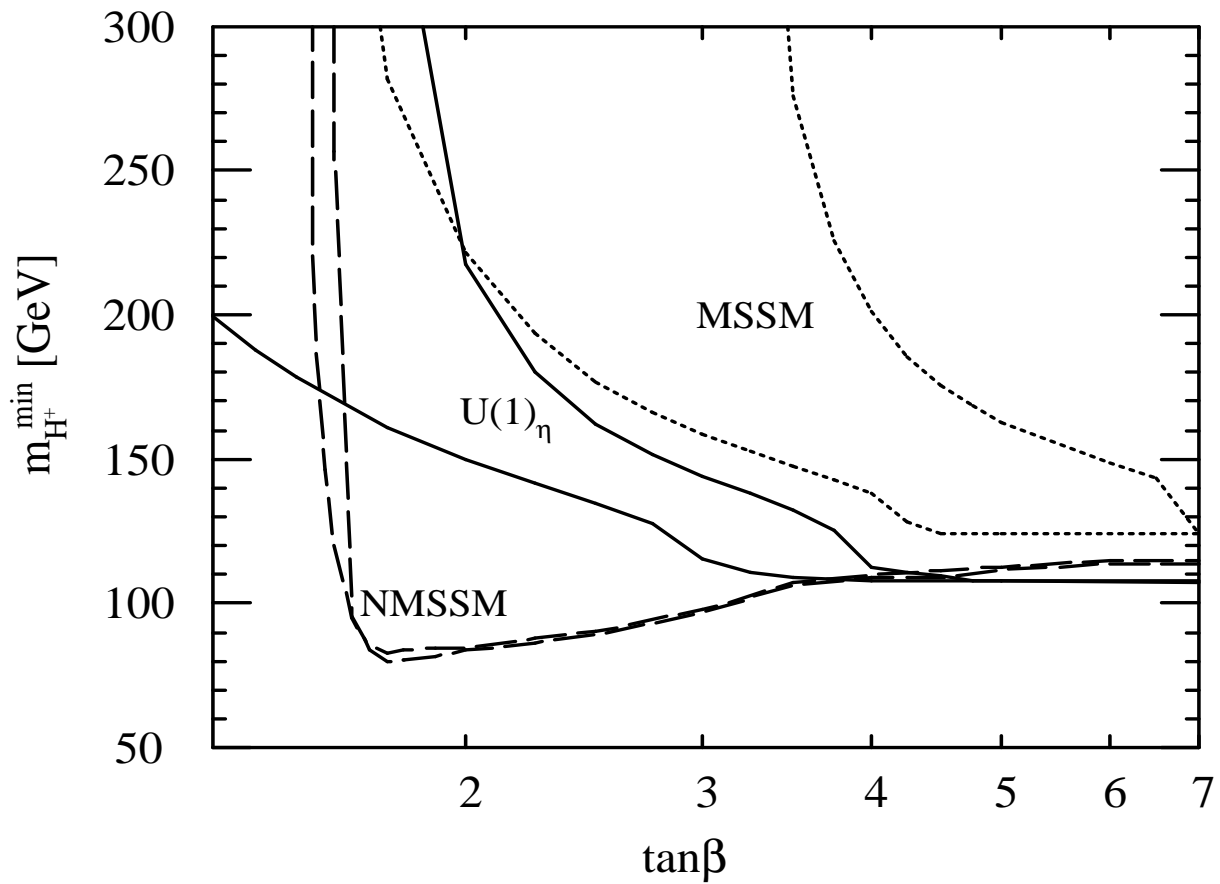,height=10cm}}
\caption
{The indirect lower bounds on \mhpl\ that could be derived if LEP fails
to find a signal for neutral Higgs bosons even after completing its run
at the projected ultimate energy of $\sqrt{s} = 200$ GeV. The notation is
as in Fig.~1}

\end{figure}

\end{document}